\documentclass[%
 reprint,
 amsmath,amssymb,groupedaddress,superscriptaddress,
 prl,longbibliography,floatfix,
]{revtex4-2}

\usepackage[T1]{fontenc}
\usepackage{lmodern}
\usepackage{graphicx}%
\usepackage{dcolumn}
\usepackage{bm}%
\usepackage[dvipsnames]{xcolor}
\usepackage{amsmath}

\usepackage{upgreek}
\usepackage{chemformula}
\usepackage{soul}
\usepackage[colorlinks,citecolor=black,linkcolor=black,urlcolor=black]{hyperref}
\usepackage{dsfont}

\hypersetup{
    colorlinks,
    linkcolor= [rgb]{0.1804,0.1882,0.5725},
    citecolor=[rgb]{0.1804,0.1882,0.5725},
    urlcolor=[rgb]{0.1804,0.1882,0.5725}
}

\usepackage{ragged2e}

\usepackage{caption}
\DeclareCaptionFormat{my}{{\bf#1#2} \justifying #3}
\begin{document}
\title{Electrically tunable Berry curvature and strong light-matter coupling in birefringent perovskite microcavities at room temperature}

\author{K.\,\L{}empicka-Mirek}
\author{M.\,Kr\'ol}
\affiliation{Institute of Experimental Physics, Faculty of Physics, University of Warsaw, ul.~Pasteura 5, PL-02-093 Warsaw, Poland}
\author{H.\,Sigurdsson}
\affiliation{Science Institute, University of Iceland, Dunhagi 3, IS-107, Reykjavik, Iceland}
\affiliation{Department of Physics and Astronomy, University of Southampton, Southampton SO17 1BJ, UK}
\author{A.\,Wincukiewicz}
\affiliation{Institute of Experimental Physics, Faculty of Physics, University of Warsaw, ul.~Pasteura 5, PL-02-093 Warsaw, Poland}
\author{P.\,Morawiak}
\affiliation{Institute of Applied Physics, Military University of Technology, Warsaw, Poland}
\author{R.\,Mazur}
\affiliation{Institute of Applied Physics, Military University of Technology, Warsaw, Poland}
\author{M.\,Muszy\'nski}
\affiliation{Institute of Experimental Physics, Faculty of Physics, University of Warsaw, ul.~Pasteura 5, PL-02-093 Warsaw, Poland}
\author{W.\,Piecek}
\affiliation{Institute of Applied Physics, Military University of Technology, Warsaw, Poland}
\author{P.\,Kula}
\affiliation{Institute of Chemistry, Military University of Technology, Warsaw, Poland}
\author{T.\,Stefaniuk}
\affiliation{Institute of Geophysics, Faculty of Physics, University of Warsaw, ul.~Pasteura 5, PL-02-093 Warsaw, Poland}
\author{M.\,Kami\'nska}
\affiliation{Institute of Experimental Physics, Faculty of Physics, University of Warsaw, ul.~Pasteura 5, PL-02-093 Warsaw, Poland}
\author{L.\,De\,Marco}
\affiliation{CNR NANOTEC, Institute of Nanotechnology, Via Monteroni, 73100 Lecce, Italy}
\author{P.\,G.\,Lagoudakis}
\affiliation{Hybrid Photonics Laboratory, Skolkovo Institute of Science and Technology, Territory of Innovation Center Skolkovo, 6 Bolshoy Boulevard 30, building 1, 121205 Moscow, Russia}
\affiliation{Department of Physics and Astronomy, University of Southampton, Southampton SO17 1BJ, UK}
\author{D.\,Ballarini}
\affiliation{CNR NANOTEC, Institute of Nanotechnology, Via Monteroni, 73100 Lecce, Italy}
\author{D.\,Sanvitto}
\affiliation{CNR NANOTEC, Institute of Nanotechnology, Via Monteroni, 73100 Lecce, Italy}
\author{J.\,Szczytko}
\author{B.\,Pi\k{e}tka}
\email{Barbara.Pietka@fuw.edu.pl}
\affiliation{Institute of Experimental Physics, Faculty of Physics, University of Warsaw, ul.~Pasteura 5, PL-02-093 Warsaw, Poland}

\begin{abstract} 
The field of spinoptronics is underpinned by good control over photonic spin-orbit coupling in devices that possess strong optical nonlinearities. Such devices might hold the key to a new era of optoelectronics where momentum and polarization degrees-of-freedom of light are interwoven and interfaced with electronics. However, manipulating photons through electrical means is a daunting task given their charge neutrality and requires complex electro-optic modulation of their medium. In this work, we present electrically tunable microcavity exciton-polariton resonances in a Rashba-Dresselhaus spin-orbit coupling field at room temperature. We show that a combination of different spin orbit coupling fields and the reduced cavity symmetry leads to tunable formation of Berry curvature, the hallmark of quantum geometrical effects. For this, we have implemented a novel architecture of a hybrid photonic structure with a two-dimensional perovskite layer incorporated into a microcavity filled with nematic liquid crystal. Our work interfaces spinoptronic devices with electronics by combining electrical control over both the strong light-matter coupling conditions and artificial gauge fields.
\end{abstract}

\maketitle

There has been surging interest of the condensed matter and solid state communities in generating artificial gauge fields across various platforms as means to describe particle properties~\cite{Aidelsburger_CRP2018} such as cold atoms~\cite{Dalibard_RMP2011, Eckardt_RevModPhys2017, Galitski_PhysTod2019, Li_LScAppl2022}, photonic materials~\cite{Fang_NatPhot2012, Lumer_NatPhot2019}, acoustics~\cite{Xiao_NatPhys2015}, mechanical systems~\cite{Abbaszadeh_PRL2017}, and exciton-polariton cavities~\cite{Gao_Nature2015, Gianfrate_Nature2020, Polimeno_Optica2021}. Gauge fields play an important role in topological properties of matter~\cite{HasanKane_RevModPhys2010, Ozawa_RevModPhys2019} and can be intimately linked~\cite{Simon_PRL1983} to a fundamental band property known as Berry curvature~\cite{berry1984quantal} quantifying the topological invariants of the system. The Berry curvature gives rise to an anomalous velocity term in a wavepacket's motion responsible for the Hall current~\cite{karplus1954hall} and also the quantum Hall effect~\cite{thouless1982quantized}, with important implications in electronic transport~\cite{Xiao_RMP2010}. 

In particular, artificial non-Abelian gauge potentials give rise to effective spin orbit coupling (SOC) of particles which has seen a lot of investigation recently in optics~\cite{Bliokh2015, Chen_NatComm2019, Yang_Science2019, Polimeno_Optica2021}. SOC forms an important ingredient in the field of spintronics~\cite{Zutic_RMP2004} and its optical analogue spinoptronics based on cavity exciton-polaritons~\cite{Liew_PhysE2011, Sanvitto_NatMat2016} seeking to exploit the mixture of internal (spin or polarization) and external (momentum) degrees of freedom for information processing. Exciton-polaritons (from here on just "polaritons") arise in the strong light-matter coupling regime as mixed states of microcavity photons and excitons~\cite{Carusotto_RMP2013}. They combine small effective photonic mass ($\sim10^{-5}$ of the electron mass) with strong nonlinear effects~\cite{StJean_NatPhoton2017} and sensitivity to external fields provided by their excitonic matter component. Moreover, they present an unique opportunity over photonic systems to study nontrivial band geometry with formation of Berry curvature~\cite{Gianfrate_Nature2020, Polimeno_NatNanotech2021} and topological effects~\cite{Karzig_PRX2015, Nalitov_PRL2015, Solnyshkov_OME2021, Klembt_Nature2018, Su2021}.

The most well known photonic SOC in microcavities is the splitting of transverse electric and transverse magnetic (i.e., TE-TM splitting) cavity photon modes~\cite{Panzarini_PRB1999}. This leads to a double winding effective magnetic field in the cavity plane which grows quadratically in the photon in-plane momentum~\cite{Leyder_NatPhys2007}. Recently, photonic analogue of the electronic Dresselhaus~\cite{Whittaker_NatPhot2020}, 
and Rashba-Dresselhaus~\cite{Rechcinska_Science2019,Ren_LPR2022} SOCs have been realized in microcavities. For the case of cavities that host both Rashba-Dresselhaus (RD) SOC and TE-TM splitting it has been shown theoretically that reducing the cavity symmetry could form local concentrations of Berry curvature~\cite{Kokhanchik_PRB2021} without the need to break time reversal symmetry through external magnetic fields acting on the excitonic component of polaritons \cite{Gutierrez_PRL2018, Bleu_PRB2018, Gianfrate_Nature2020} or nonzero optical activity for the photons \cite{Ren_NatCommun2021,Polimeno_NatNanotech2021}.
\begin{figure*}
		\centering
		\includegraphics[width=.99\textwidth]{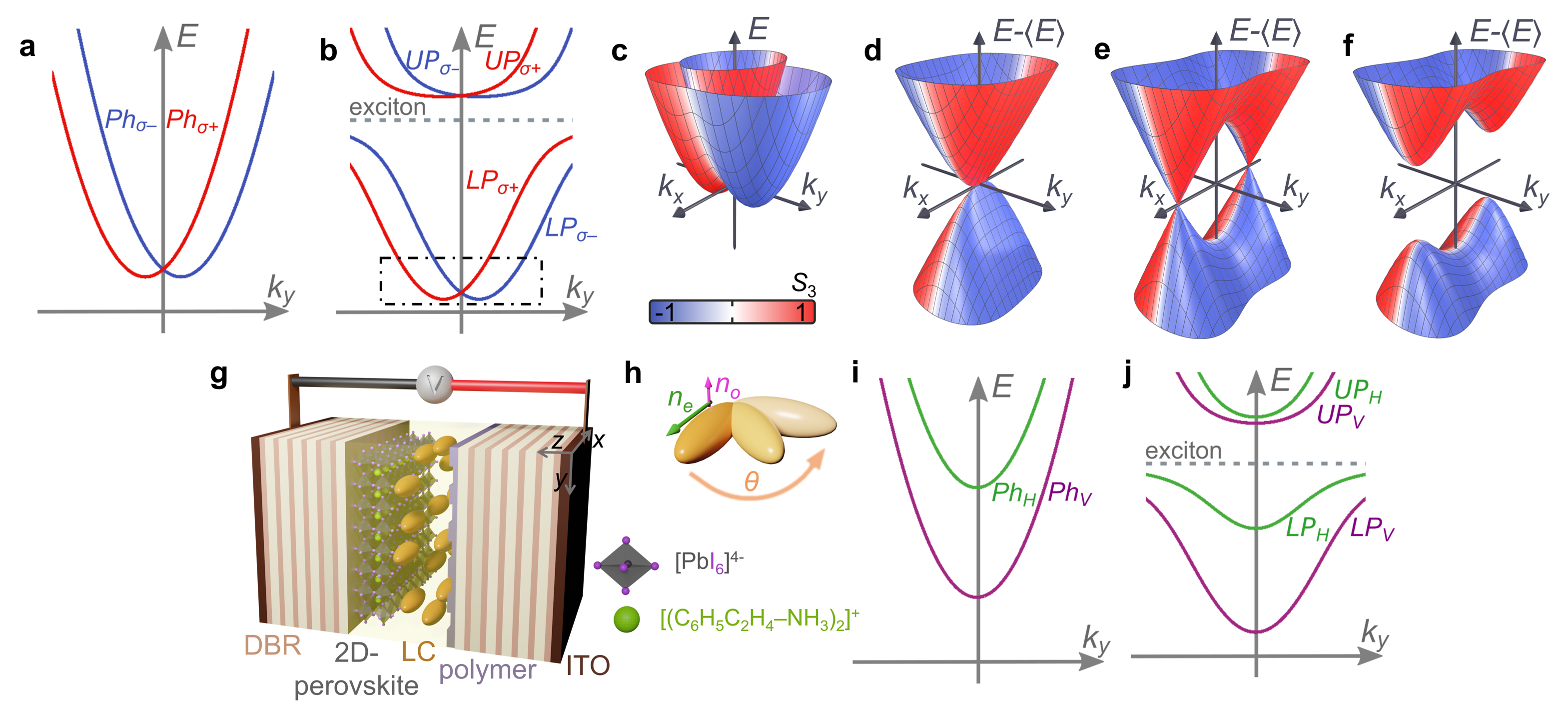}
		\caption {\textbf{Liquid crystal microcavity with 2D perovskite.} Schematic dispersion relation of \textbf{a} bare cavity photon modes in the RD SOC regime and \textbf{b} in the strong light-matter coupling regime. \textbf{c} Dispersion relation for bottom of lower polariton branch (the region marked by a dashed square in panel \textbf{b}). \textbf{d} Same as \textbf{c} with the dispersion subtracted by its mean value $\langle E \rangle$ to more clearly show the intersection points. Energy of the modes for \textbf{e} positive H-V splitting and \textbf{f} with broken inversion symmetry. \textbf{g} Schematic representation of the LC microcavity with the 2D-hybrid organic-inorganic perovskite layer and LC molecules oriented parallel to the $x$-$y$ plane at zero voltage. \textbf{h}~Rotation of a LC molecule under applied voltage with refractive indices ($n_\text{o}$, $n_\text{e}$) and angle $\theta$ in the $z$-$x$ plane. We also show the bare photon \textbf{i} and polariton \textbf{j} dispersions in the linear horizontal-vertical (H-V) polarization basis in the absence of voltage to highlight their splitting.}
\label{rys:_FIG1_v4}
\end{figure*}

In this work we present a method to electrically tune photonic Berry curvature in the strong light-matter coupling regime. Our optical system is composed of a liquid crystal (LC) cavity where both RD SOC and TE-TM splitting effects of the cavity photons are inherited in the emerging exciton-polariton modes due to an additional cavity embedded perovskite layer. The strongly bound perovskite excitons allow observation of strong coupling at room temperature with high quantum yield and nonlinear effects up to four orders of magnitude higher than in other photonic systems~\cite{Polimeno_Optica2021,Wu2021}, and thus present a good material to be interfaced with LCs. Due to the high birefringence and electric permittivity anisotropy of LCs, making them sensitive to external electric fields, we achieve unprecedented electric control over an emerging polaritonic Berry curvature through the specially synthesized photonic SOC and reduced cavity symmetry.

A cross section of the photonic RD dispersion in the cavity ($k_x,k_y$) plane is schematically shown in Fig.\,\ref{rys:_FIG1_v4}a depicting two opposite circularly polarized valleys in analogy with spin 1/2 systems. When the photons become strongly coupled with the perovskite excitons a characteristic anticrossing behaviour occurs shown in Fig.\,\ref{rys:_FIG1_v4}b where the low energy polariton modes (within the dot-dashed square box) adopt approximately the same dispersion as that of the photons (Fig.\,\ref{rys:_FIG1_v4}a). A surface plot of the polariton dispersion at low momenta is shown in Fig.\,\ref{rys:_FIG1_v4}c. A single degeneracy point at normal incidence ($\mathbf{k}=0$) is highlighted in the RD polariton dispersion (Fig.\,\ref{rys:_FIG1_v4}d) where we have subtracted the dispersion from its mean $\left\langle E \right\rangle$. When TE-TM splitting and an additional uniform in-plane effective magnetic field $\mathbf{B}_x = \Delta_{HV} \mathbf{\hat{x}}$ are present the RD degeneracy point morphs into two Dirac cones with degeneracy points known as diabolical points (Fig.\,\ref{rys:_FIG1_v4}e)~\cite{Su2021, Gianfrate_Nature2020, Ren_NatCommun2021, Spencer_SciAdv2021, Bergholtz_RevModPhys2021}. The rapidly whirling RD SOC field around these points leads to a topologically trivial gap opening (Fig.\,\ref{rys:_FIG1_v4}f) when an additional perpendicular $\mathbf{B}_y = \Delta_{AD} \mathbf{\hat{y}}$ effective magnetic field is introduced to break the inversion symmetry of the system with subsequent formation of a Berry curvature dipole (as shown later in the text).
    \begin{figure*}
		\centering
		\includegraphics[width=.99\textwidth]{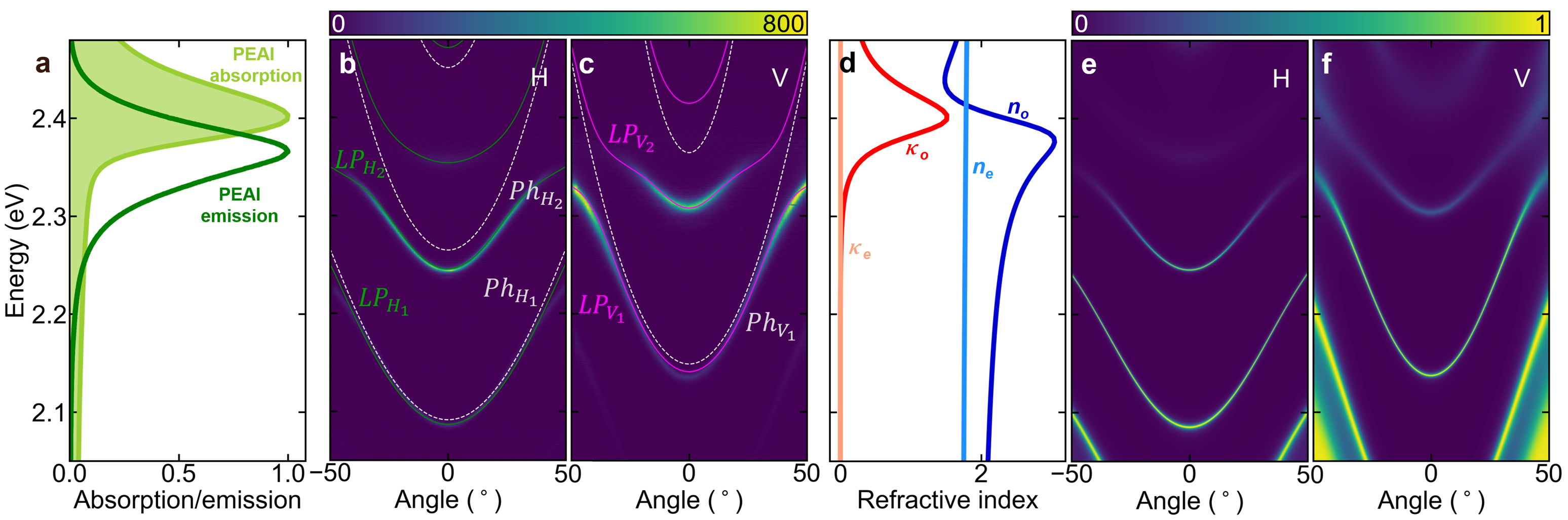}
		\caption{\textbf{Strong light-matter coupling regime in liquid crystal microcavity with 2D perovskite.} \textbf{a}~Normalized absorption (light green line) and emission (dark green line) spectra of a polycrystalline 2D-perovskite.
		\mbox{\textbf{b--c}}\,Angle-resolved photoluminescence spectra showing the strong-coupling dispersion in horizontal (H) and vertical (V) polarization at zero voltage ($\theta=0$). 
		White dashed curves indicate calculated bare photon dispersion, and green and magenta curves the calculated polariton dispersion (see Methods). \textbf{d} Real $n_{\text{o,e}}$ and imaginary $\kappa_{\text{o,e}}$ parts of the ordinary and extraordinary refractive indices for a thin polycrystaline 2D-perovskite layer.
		\textbf{e--f} Berreman simulations corresponding to panels \textbf{b--c} (see Methods).}
		\label{rys:_FIG2_v1}
	\end{figure*}

\subsection{Strong light-matter coupling}

The microcavity consists of two distributed Bragg reflectors (DBRs) facing each other with an embedded 60\,nm thick polycrystalline 2D phenylethylammonium lead iodide (PEAI) perovskite (C$_{6}$H$_{5}$C$_{2}$H$_{4}$NH$_{3}$)$_{2}$PbI$_{4}$ (hereafter termed: PEPI), shown schematically in Fig.~\ref{rys:_FIG1_v4}g.
The perovskite polycrystalline thin film was prepared on one of the DBR inner sides using a spin-coating method~\cite{Lempicka2019}. The cavity structure was designed to enhance the photonic field at the position of the perovskite layer. The cavity was also filled with a highly birefringent nematic LC which acts as an uniaxial medium with ordinary $n_\text{o}$ and extraordinary $n_\text{e}$ refractive indices ($\Delta n = n_\text{e}-n_\text{o}=0.4$ \cite{Miszczyk_LC2018}). The voltage applied to transparent electrodes (made of Indium Tin Oxide - ITO) rotates the molecular director, hence the direction of the optical axis of the LC medium, by an angle $\theta$ in the $x$-$z$ plane (see Fig.~\ref{rys:_FIG1_v4}h). This enables direct control over the cavity effective refractive indices. Throughout the paper we will refer to the Stokes parameters of the emitted cavity light (analogous to the polariton pseudospin) as $S_1=(I_H - I_V)/(I_H + I_V)$ and $S_2=(I_D - I_{AD})/(I_D + I_{AD})$ and $S_3=(I_{\sigma^+} - I_{\sigma^-})/(I_{\sigma^+} + I_{\sigma^-})$ corresponding to intensities of horizontal ($I_H$), vertical ($I_V$), diagonal ($I_D$), antidiagonal ($I_{AD}$), right-hand circular ($I_{\sigma^+}$) and left-hand circular ($I_{\sigma^-}$)  polarized light.

The alignment of the LC molecules inside the cavity at zero voltage is determined by an ordering polymer layer rubbed along the $x$ (H) direction as shown schematically in Fig.~\ref{rys:_FIG1_v4}g. Consequently, the bare photon dispersion in the linear horizontal-vertical (H-V) polarization basis shown in Fig.\,\ref{rys:_FIG1_v4}i displays strongly split bands described by the Hamiltonian:
\begin{equation}
        \hat{H}_\phi(\mathbf{k},\theta) = \frac{\hbar^2}{2}\left(\frac{k_x^2}{m_x} + \frac{k_y^2}{m_y}\right) + \hat{S}(\mathbf{k}) + \frac{\Delta_{HV}}{2} \hat{\sigma}_z,
        \label{eq:Linear2x2}
\end{equation}
where $\mathbf{k} = (k_x,k_y)^{\rm T}$ is the in-plane momentum. Here, $\hat{S}$ describes an effective photonic SOC coming from both cavity TE-TM splitting~\cite{Panzarini_PRB1999} and the LC anisotropy~\cite{Rechcinska_Science2019},
\begin{equation}
\hat{S}  = 2 \delta_{xy} k_x k_y \hat{\sigma}_x  + \left( \delta_x k_x^2 - \delta_y k_y^2 \right) \hat{\sigma}_z.
\label{eq.S}
\end{equation}
Physically, this operator introduces direction dependent modification to the effective masses of linearly polarized modes. When $\delta_{xy}=\delta_y=\delta_x$ it realizes the conventional TE-TM splitting~\cite{Kavokin_PRL2005}. Finally, $\Delta_{HV}$ describes uniform splitting between H and V polarized modes (sometimes referred as $X$-$Y$ splitting) which can be directly controlled in the experiment through the voltage applied to the ITO electrodes on the cavity, where the electric field drives the molecular director, hence the optical axis and changes the effective refractive indices for the H polarized mode. We note that all coefficients in  Eqs.~\eqref{eq:Linear2x2} and~\eqref{eq.S} depend on $\theta$ but not as strongly as $\Delta_{HV}$~\cite{Rechcinska_Science2019}. We do not consider non-Hermitian effects which are weak in our system and only relevant to the physics of exceptional points~\cite{Su2021, Liao_PRL2021, Septembre_arxiv2021} which is beyond the scope of our study.
	
When an exciton resonance, with energy $E_\chi$, from the perovskite layer is introduced to the cavity (dashed horizontal line in Fig.\,\ref{rys:_FIG1_v4}j) the linearly polarized photonic modes coherently couple to excitons at a rate defined by the Rabi energies $\Omega_{H,V}$. The strongly coupled system can be described by a coupled-oscillators model represented by a Hamiltonian:
    \begin{equation}
        \hat{H}_{SC} = 
        \begin{pmatrix}
        E_\chi & 0 \\ 0 & 0 
        \end{pmatrix} \otimes
         \mathds{1}_2  + 
        \begin{pmatrix}
        0 & 0 \\ 0 & 1
        \end{pmatrix} \otimes
        \hat{H}_\phi +
        \frac{\hat{\sigma}_x}{2} \otimes 
        \begin{pmatrix}
        \Omega_{\rm H} & 0 \\
        0 & \Omega_{\rm V}
        \end{pmatrix}
        \label{eq:SCLinear2x2}
    \end{equation}
The four polariton eigenmodes are shown in \textbf{Figure\,\ref{rys:_FIG1_v4}}j, exhibiting characteristic anticrossing behaviour, and labelled as upper (UP$_{H,V}$) and lower polaritons (LP$_{H,V}$).
\begin{figure*}
		\centering
		\includegraphics[width=.9\textwidth]{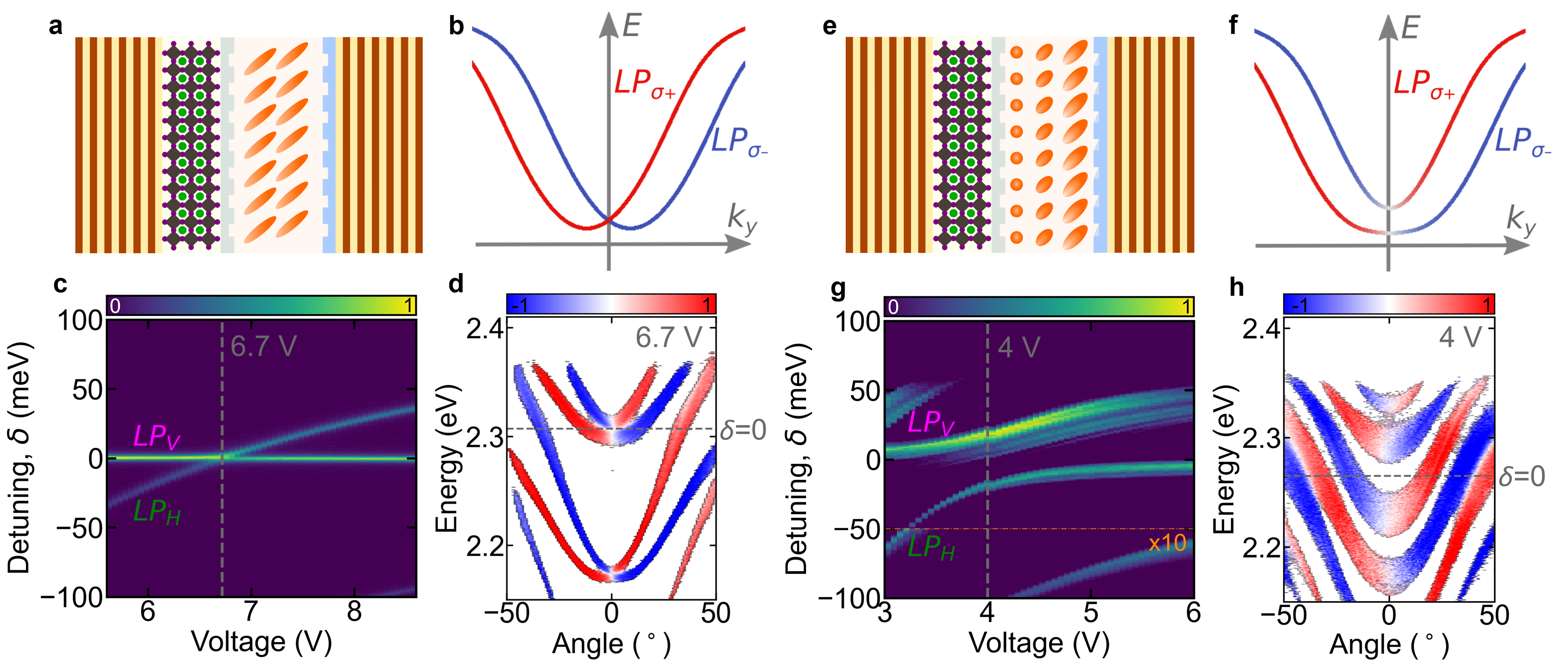}
		\caption{\textbf{Gap opening in the Rashba-Dresselhaus regime.} \textbf{a,e} Scheme of a 2D-polycrystalline perovskite microcavity with a LC having: two parallel ordering polymer layers at both sides of the cavity (\textbf{a}) and two ordering polymer layers at both sides of the cavity with 45$^{\circ}$ tilted rubbing (\textbf{e}) (see Methods). \textbf{b,f} Dispersion relation of first polariton branch in Rashba-Dresselhaus SOC regime (\textbf{b}), with the opening of the energy gap (\textbf{f}). \textbf{c,g} Measured energy resolved PL at $\mathbf{k}=0$ versus applied voltage for diagonal polarization: without (\textbf{c}) and with (\textbf{g}) gap opening. \textbf{d,h} Measured angle-resolved PL spectra in R-D regime for $S_{3}$ polarization: without (\textbf{d}) and with (\textbf{h)} gap opening.}
		\label{rys:_FIG3_v3}
\end{figure*}

The strong coupling regime in our structure is illustrated in Fig.~\ref{rys:_FIG2_v1}. The strong emission (dark green line) and absorption (light green line) spectra of a thin PEPI layer without the cavity are shown in Fig.~\ref{rys:_FIG2_v1}a. The maximum of the spectra overlap is at 2.38~eV, which corresponds to the most effective emission and reabsorption processes. The emission spectra from our cavity at room temperature at zero voltage (i.e., $\theta = 0$) are presented in Fig.~\ref{rys:_FIG2_v1}b,c evidencing strong emission from H-V polariton modes. The characteristic anti-crossing between the excitonic resonance and the cavity photon modes is visible in the emission spectra at high emission angles.

Our measurements are compared with the solutions of Eq.~\eqref{eq:SCLinear2x2} plotted with solid lines in Fig.~\ref{rys:_FIG2_v1}b,c (see Methods). The white dashed lines indicate the fitted bare photonic branches. Extracted Rabi energies for H and V polarized modes are: $\Omega_{H}=94.4$~meV and $\Omega_{V}=108.7$~meV. This difference between the Rabi energies is due to different optical paths for the two linear polarizations (difference in the LC refractive indices $n_{\rm o}$ and $n_{\rm e}$). 

We also performed numerical simulations on the optical properties of our cavity using the Berreman matrix method. For this purpose, the real and imaginary parts of the ordinary and extraordinary refractive indices $n_{\text{o,e}}$ and $\kappa_{\text{o,e}}$, respectively, were obtained from ellipsometric measurements and are presented in Fig.~\ref{rys:_FIG2_v1}d. We observe a slight birefringence of the polycrystalline perovskite in the $z$-axis direction (perpendicular to the cavity plane). The simulated angle-resolved transmission spectra are shown in Fig.~\ref{rys:_FIG2_v1}e,f. The theoretical result is fully consistent with the experiment. The higher energy modes (above 2.39\,eV) are not visible in the experimental spectra due to the strong absorption of the perovskite in this spectral range.

\subsection{Rashba-Dresselhaus polaritons}
By rotating the molecular director of the LC with applied voltage, photonic modes of different polarization and parities become mixed and form a RD SOC dispersion relation~\cite{Rechcinska_Science2019}. In this case, any effective in-plane fields are minimal and the photonic modes are circularly polarized, forming a dispersion depicting two shifted valleys of opposite circular polarization (like shown in Fig.\,\ref{rys:_FIG1_v4}a). In this regime, the photonic Hamiltonian can be written: 
\begin{equation}
\hat{H}_\phi^{RD}(\theta, \mathbf{k})    = \hat{H}_\phi(\theta, \mathbf{k}) - 2 \alpha k_y \hat{\sigma}_y
\label{eq:Rashba2x2}
\end{equation}
where $\alpha$ is the strength of the RD SOC. It is worth noting that in this low-energy regime polariton and photon lasing was recently demonstrated\cite{Li_arxiv2021,Muszynski_PRApp2021}. 

The polariton RD SOC regime is illustrated in Fig.~\ref{rys:_FIG3_v3} for two different cavity types. In Fig.~\ref{rys:_FIG3_v3}a-d we show the effects of RD SOC photonic modes coupled to the excitonic resonance. Here, the LC molecules are initially (at 0\,V) oriented by two parallel rubbing polymer layers as shown in Fig.~\ref{rys:_FIG3_v3}a. 
The degeneracy point in the RD dispersion (see Fig.~\ref{rys:_FIG3_v3}b at $\mathbf{k}=0$) is sensitive to H-V splitting that can be directly controlled through the applied voltage [i.e., $\Delta_{HV}(\theta)$]. We measured the cavity PL as a function of applied voltage at $\mathbf{k}=0$, shown in Figure~\ref{rys:_FIG3_v3}c, and observed the crossing of two polariton modes at around 6.7\,V. This crossing point corresponds to the degeneracy point in the RD polariton dispersion, also evidenced through the measured $S_3$ component shown in Fig.~\ref{rys:_FIG3_v3}d.
\begin{figure*}
		\centering
		\includegraphics{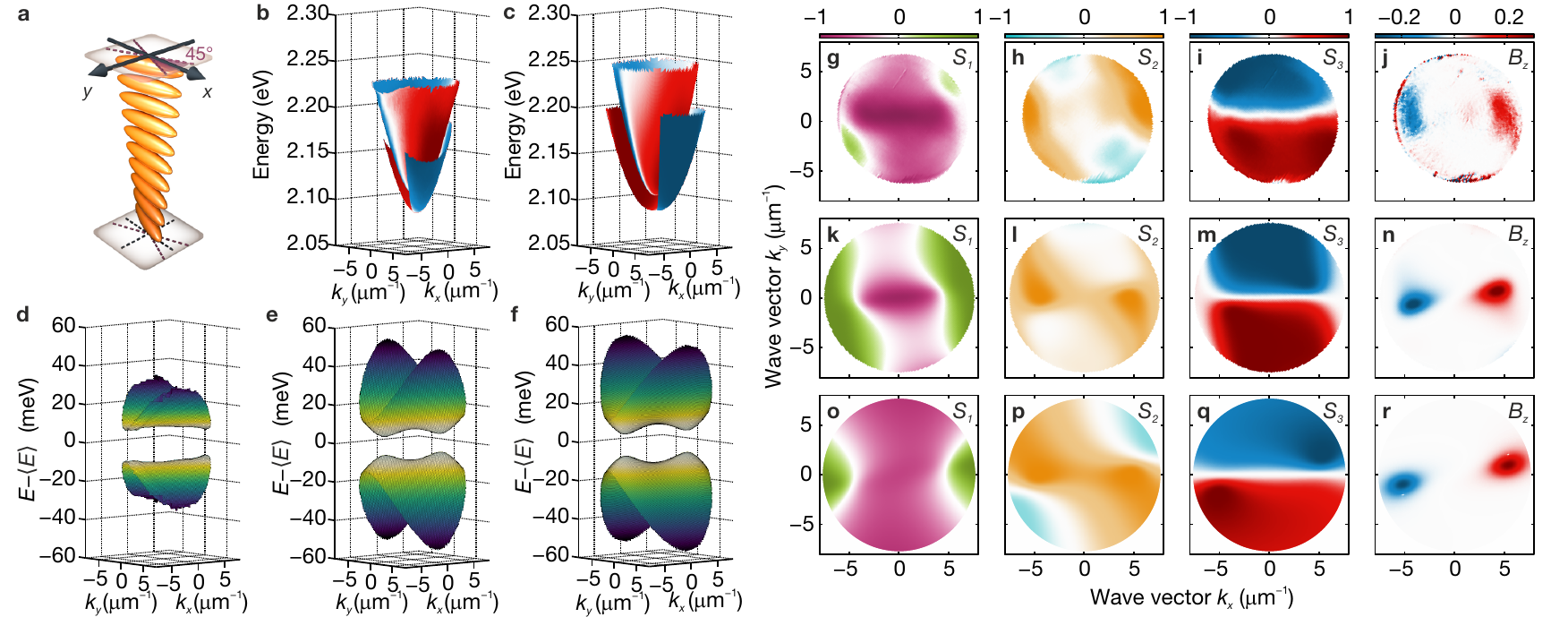}
		\caption{\textbf{Nonzero Berry curvature in LC cavity with 2D-perovskite.} \textbf{a} Scheme of orientation of LC molecules inside the cavity in twisted nematic configuration. Dispersion of polaritonic modes obtained by \textbf{b} experiment and \textbf{c} Berreman matrix simulations. The pseudocolor scale denotes the $S_3$ component.  \textbf{d} Measured polariton bands subtracted from their mean energy with corresponding \textbf{e} Berreman matrix simulations and \textbf{f} solutions of the analytical model. \textbf{g--j} Experimental polarization and resulting Berry curvature dipole of the lower energy band compared with \textbf{k--n} results of the Berreman simulations and \textbf{o--r} analytical two-level model.}
		\label{rys:FIG5}
\end{figure*}

\subsection{Berry curvature dipole}
Synthesizing polaritonic Berry curvature in optical cavities is a challenging task since it requires breaking of time-reversal symmetry  through external magnetic fields acting on the excitonic component \cite{Gutierrez_PRL2018, Bleu_PRB2018, Gianfrate_Nature2020, Polimeno_NatNanotech2021}, or nonzero optical activity for the photons \cite{Ren_NatCommun2021}. Such chiral terms open a gap at the diabolical points in the polariton dispersion, corresponding to a Dirac cone intersection~\cite{Tercas_PRL2015,Polimeno_Optica2021, Bieganska_PRL2021,Spencer_SciAdv2021}, where the TE-TM SOC field from $\hat{S}$ and the uniform field $\Delta_{HV}$ cancel each other out. Indeed, with $\alpha=0$ the polariton modes are linearly polarized with a rapidly changing in-plane SOC field winding around the Dirac cones (i.e., the polariton pseudospin winds around the $S_1$--$S_2$ plane when going around the diabolical point~\cite{Gianfrate_Nature2020,Polimeno_Optica2021}). It is worth noting that when non-Hermitian terms are present these diabolical points become exceptional points which expand to degeneracy lines (Fermi arcs) when polarization-dependent losses are present~\cite{Richter_PRA2017, Richter_PRL2019, Su2021, Septembre_arxiv2021}. 

Interestingly, when RD SOC is present it introduces and additional rapidly varying field around the Dirac cones manifested in the strong $S_3$ component in the polariton dispersion. Practically, the diabolical points can only appear along the ($k_x,k_y=0$) axis where the RD field vanishes (see Fig.~\ref{rys:_FIG1_v4}e). Therefore, the RD field will only affect the SOC field around these points but it cannot open the gap. However, because of the combined TE-TM and RD SOC fields the polariton pseudospin no longer winds just in the $S_1$--$S_2$ plane when going around the diabolical points but also around the $S_1$--$S_3$ plane. This means that an effective magnetic field in the $y$-direction corresponding to a finite $S_2$ pseudospin component will break inversion symmetry at the diabolical points and open the gap. Physically, such an effective magnetic field corresponds to splitting between the diagonal (D) and antidiagonal (AD) modes written,
\begin{equation}
\hat{H}_\phi^{RD}(\theta, \mathbf{k})    = \hat{H}_\phi(\theta, \mathbf{k}) - 2 \alpha k_y \hat{\sigma}_y + \frac{\Delta_{AD}}{2} \hat{\sigma}_x.
\label{eq:Rashba2x2_2}
\end{equation}
Indeed, setting $k_y=0$ it can be shown that the presence of the $\Delta_{AD}$ breaks parity inversion symmetry $\hat{\sigma}_z \hat{H}^{RD}_\phi(-k_x,0) \hat{\sigma}_z  \neq  \hat{H}^{RD}_\phi(k_x,0)$ where $\hat{\sigma}_z$ flips the photon circular polarization in the H-V basis. This leads to opening of the gap with subsequent formation of local dispersion relation describing massive Dirac particles  (see Fig.~\ref{rys:_FIG1_v4}f).
\begin{figure*}
		\centering
		\includegraphics{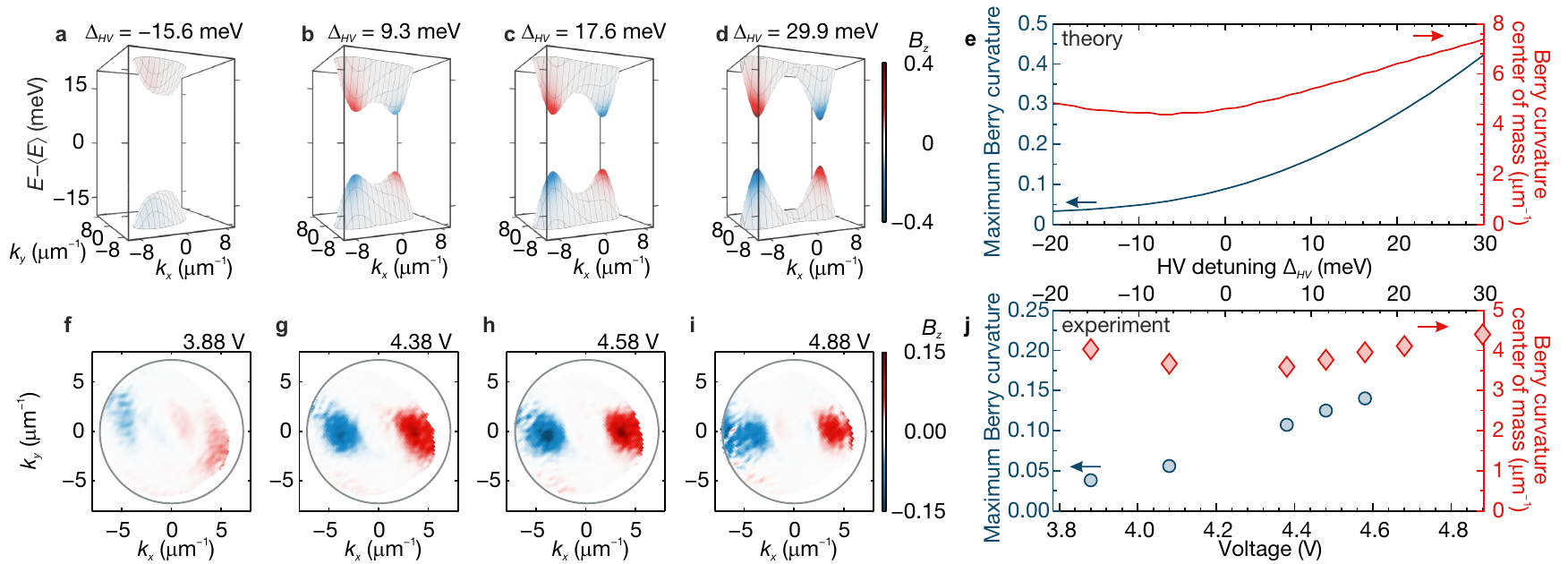}
		\caption{\textbf{Electrical tuning of polariton Berry curvature}. \textbf{a}--\textbf{d} Calculated dispersion of the cavity modes (energies subtracted by their mean) with $B_z$ marked by a pseudocolor scale from the analytical model for increasing H-V detuning $\Delta_{HV}$. \textbf{e} Resulting maximum value and center of mass position of the $B_z$ in reciprocal space for varying H-V detuning. \textbf{f}--\textbf{i} Experimental $B_z$ distribution for lower energy mode with increasing voltage. \textbf{j}~ Experimentally measured maximum value of $B_z$ and $B_z$ center of mass position depending on applied voltage.}
		\label{rys:FIG6}
\end{figure*}

For this purpose, we have prepared a second sample with rubbing at the perovskite layer oriented at an angle of 45$^{\circ}$ with respect to the rubbing at the DBR surface, Fig.~\ref{rys:_FIG3_v3}e. This results in a twisted nematic LC structure (as illustrated in Fig.~\ref{rys:FIG5}a) which introduces a deterministic D-AD splitting. The splitting is confirmed in Fig.~\ref{rys:_FIG3_v3}g where the PL spectrum at $\mathbf{k}=0$ as a function of voltage shows a clear anticrossing. Fig.~\ref{rys:_FIG3_v3}h shows the circular polarization degree of the corresponding angle-resolved dispersion and Fig.~\ref{rys:_FIG3_v3}f the calculated dispersion. 

The concave energy relation indicating the location of the massive Dirac cones with a clear splitting is illustrated in Fig.~\ref{rys:FIG5}b-f. The Stokes (pseudospin) parameters in Fig.~\ref{rys:FIG5}g-i show that at the minimum of energy-gap the $S_1$ and $S_3$ polarizations of the bands whirl around two diagonally polarized points located along $k_x$. This evidences that the splitting between D and AD polarized modes in the twisted nematic configuration has opened a gap at these points. More importantly, these points are accompanied with strong concentration of polaritonic Berry curvature as measured and calculated in Fig.~\ref{rys:FIG5}j,n,r. As time reversal symmetry is conserved in the system integral over the the Berry curvature within a single band is equal to zero, and for the second band the Berry curvature is expected to be of the opposite sign. We note that at $\mathbf{k} = 0$ in Fig.~\ref{rys:FIG5}g the mode is predominantly V polarized due to the applied voltage causing some additional splitting $\Delta_{HV}$. All of our results are in qualitative agreement with both our analytical model of Eq.~\eqref{eq:Rashba2x2_2} and Berreman simulations. Results of the analytical model in Fig.\,\ref{rys:FIG5}f were fitted to the dispersion relation obtained from Berreman matrix model (see Methods).

Lastly, the observed polariton Berry curvature can be electrically tuned through the parameter $\Delta_{HV}$ which is directly controlled by amplitude of external voltage applied to the cavity. Fig.~\ref{rys:FIG6}a--e presents calculations based on Eq.~\eqref{eq:Rashba2x2} for varying detuning between the H-V modes. The calculated energies of the two bands are subtracted from their mean value with the Berry curvature $B_z$ marked by a pseudocolor scale. The Berry curvature increases with H-V detuning as the dispersion relation changes to two gapped Dirac cones exhibiting a pronounced Berry curvature. Dependence of the maximum value of Berry curvature and the $B_z$ center of mass position in reciprocal space is summarized in Fig.~\ref{rys:FIG6}e, which shows that the momentum position of the Berry curvature maximum can be controlled by $\Delta_{HV}$ detuning and can be effectively switched off for negative detunings.  

Those theoretical predictions can be confirmed experimentally for varying voltage applied to the cavity which directly controls $\Delta_{HV}$ detuning. Fig.\,\ref{rys:FIG6}f--i presents Berry curvature extracted from polarization-resolved transmission measurement at different voltages applied to the ITO electrodes of the cavity. At 3.88\,V (Fig.\,\ref{rys:FIG6}f) corresponding to negative detuning observed value of the Berry curvatures is low, but rapidly increases for higher voltages when $\Delta_{HV}$ detuning becomes positive (Fig.\,\ref{rys:FIG6}g). Further increase of the applied voltage shifts the position of the $B_z$ maxima to higher momenta in reciprocal space up to the limits of the numerical aperture in the experiment (Fig.\,\ref{rys:FIG6}h,i). Experimental position and value of the polaritonic Berry curvature summarised in Fig.\,\ref{rys:FIG6}j are in good agreement with theoretical predictions of Fig.\,\ref{rys:FIG6}e and clearly demonstrate electric tunability of the band geometry in our system. The polaritonic Berry curvature in our system is tunable in a continuous manner with the amplitude of external voltage, rather than magnetic field or temperature \cite{Polimeno_NatNanotech2021}.

\section*{Conclusion}
In conclusion, we have presented a novel architecture for a photonic cavity heterostructure with a 2D-perovskite incorporated in a highly birefringent medium. The medium is a nematic liquid crystal that can be arranged in various ways into the cavity through proper preparation of embedded ordering layers. We firstly achieve electrically tunable strong light-matter coupling between the perovskite excitons and cavity photons as a functing of applied external electric field regime at room temperature with Rabi splitting in the order of $\sim100$~meV. We demonstrate how the extreme birefringence coming from the liquid crystal results in exciton-polariton modes following a Rashba-Dresselhaus spin-orbit-coupling dispersion relation which opens exciting perspectives on designing nonlinear valley-optronic devices. 

Second, the reduced symmetry of the cavity structure is found to open a gap at polaritonic diabolical points corresponding to Dirac cones in the cavity dispersion relation. This allows us to engineer local concentrations of Berry curvature that can be controlled through applied external voltage which has only been possible using external magnetic fields or temperature variations~\cite{Polimeno_NatNanotech2021}. The importance of the Berry curvature dipole to the quantum nonlinear Hall effect~\cite{Sodemann_PRL2015} and rectification in polar semiconductors~\cite{Ideue_NatPhys2017} was recently stressed in a study demonstrating its electric tunability in a monolayer WTe$_2$~\cite{Xu_NatPhys2018}. Our work therefore takes steps towards interfacing polariton spinoptronic technologies with electronics at room temperature where nontrivial geometrical properties of the polariton dispersion can be easily tuned through applied voltage. We note that both electrically pumped polariton lasers~\cite{Schneider_Nature2013} and electro-optic modulated polariton topological lasers~\cite{Gagel_ACSPho2022} are already possible at cryogenic temperatures. Moreover, with already exciting work aimed at Rashba-Dresselhaus polariton condensation~\cite{Li_arxiv2021} we expect that our platform can serve as a testbed for driven-dissipative quantum matter in unconventional artificial gauge fields.

%

\clearpage
\noindent {\bf\Large Methods} \\
    
\noindent{\bf Preparation of liquid crystal microcavity with 2D-perovskite.}
     The studied microcavity structure consists of two distributed Bragg reflectors (DBRs) made of 6 SiO$_{2}$/TiO$_{2}$ (SiO$_{2}$ top layer) pairs with maximum reflectance at 520~nm. DBRs grown on glass substrates with ITO electrodes. The surface of the dielectric mirror was cleaned with isopropanol and acetone to remove organic residues from the surface. It was then activated in an oxygen plasma, thanks to which the 2D-perovskite crystallization precursor solution adhered better to the substrate. Due to the activation of the DBRs surface, the 2D-perovskite crystallization precursor solution adheres better to the substrate. The precursor solution was prepared in a glove box under an argon atmosphere as follows: phenethylammonium iodide (PEAI) was mixed with lead iodide (PbI$_{2}$) in a stoichiometric 2:1 molar ratio and dissolved in N,N-dimethylformamide (DMF) (mass percent: 10\%). The solution was stirred for 4\,h at 50\,$^{\circ}$C. All reagents were purchased from Sigma Aldrich. 40\,$\mu$l of solution containing PEAI and PbI$_{2}$ was spin-coated in air on a dielectric mirror with rotation speed of 2,000 rpm for 30 sec. Approximately 60~nm thick 2D-layered polycrystalline lead iodide perovskite [(C$_{6}$H$_{5}$(CH$_{2}$)$_{2}$NH$_{3}$)$_{2}$PbI$_{4}$] was obtained upon solvent evaporation on the hotplate for 1 minute at 100\,$^{\circ}$C. 
    
    There were two types of microcavities as shown in (Fig.\,\ref{rys:_FIG3_v3}). For these samples, the preparation of the dielectric mirror coated with 2D-polycrystalline perovskite was the same, as described above. In the case of the first sample (\text{Fig.\,\ref{rys:_FIG3_v3}}a), the bottom dielectric mirror with perovskite was spin-coated with a thin, protective layer (50~nm thick) of PMMA with a 3\% methoxybenzene solution. The upper dielectric mirror was covered with a structured polymer orienting layer of SE-130. The space between the DBRs is filled with highly birefringent liquid crystal of $\Delta n=0.4$. Both mirrors were rubbed to obtain homogeneous LC orientation within the cavity. The total thickness of this liquid crystal cavity with was approximately 3~$\mu$m. The second sample type was similar in design to the first one, but the PMMA layer on the dielectric mirror was arranged at an angle of 45$^{\circ}$ (Fig.\,\ref{rys:_FIG3_v3}e,g,h, Fig.\,\ref{rys:FIG5}) or 30$^{\circ}$ (Fig.\,\ref{rys:FIG6}) to the rubbed layer of SE-130 polymer. The thickness of those cavities was approximately 2.5~$\mu$m. The different thickness of the presented three perovskite liquid crystal cavities gives a different range of liquid crystal voltage tuning to the Rashba-Dresselhaus regime (Fig.~\ref{rys:_FIG3_v3}c,g,  Fig.\,S7, Fig.\,S8).

    \medskip
    
\begin{figure}[htp]
    		\centering
    		\includegraphics[width=0.5\textwidth]{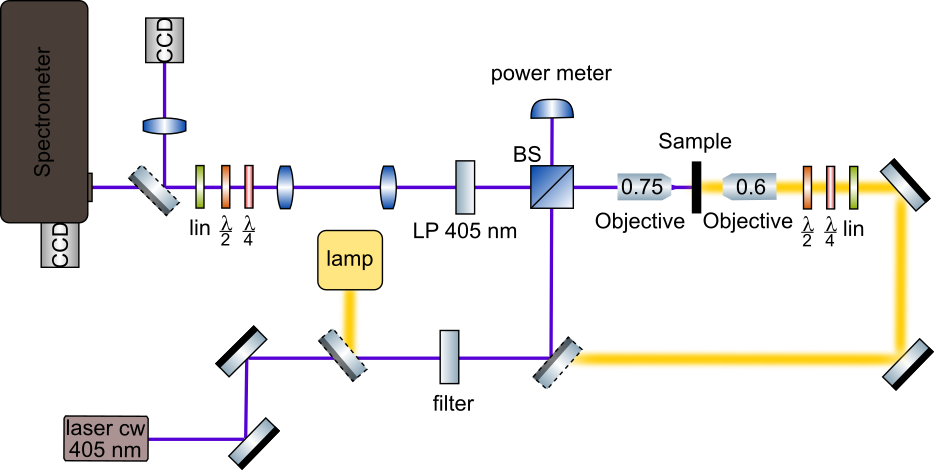}
    		\caption{\text{Scheme of experimental set-up.}}
    		\label{rys:set_up}
          \end{figure}

\noindent{\bf Optical measurements.}
          \text{Figure~\ref{rys:set_up}} shows the scheme of the experimental setup used for the optical measurements in both configuration: reflectance and transmission. The measurements were performed at room temperature using a continuous wave laser at 405~nm for photoluminescence (PL) excitation and a white light lamp for the reflectance and transmission spectra. PL measurements were realized in the reflection configuration.
          
\medskip

\noindent{\bf Absorbance measurements.}
      Absorption measurements of the PEPI perovskite on the glass substrate were performed using a CARY 5000 UV-Vis-NIR spectrometer. The absorption spectrum is presented in Fig.~\ref{rys:_FIG2_v1}a.
      
      \medskip
      
\noindent{\bf Ellipsometry measurements.}
      The refractive index of polycrystalline PEPI perovskite (presented in Fig.~\ref{rys:_FIG2_v1}d) was estimated based on ellipsometric data from RC2 ellipsometer (J.A. Woollam Co) in 450-1690\,nm spectral range. To increase the sensitivity of the measurement and retrieve anisotropy more accurately two different types of samples were prepared. In the first set of samples, PEPI was spin-coated on the silicon wafer with a 618\,nm thick layer of thermal oxide to utilize the interference enhancement effect. Then all 16 Mueller matrix (MM) elements in reflection were acquired for angles of incidence from 55$^\circ$ to 70$^\circ$ by 5$^\circ$. In the second case, the PEPI layer was spin-coated on a 1 mm thick transparent fused silica substrate. For this set of samples, transmission ellipsometry and transmission intensity data for illumination angles ranging from 0$^\circ$ to 40$^\circ$ by 5$^\circ$ were acquired additionally. Such a combination helps to reduce the correlation between the model parameters, leading to a unique solution. Almost negligible values of off-diagonal terms and off-diagonal blocks in the MM data collected in the reflection mode reveal no cross-polarization between $p$- and $s$- states and indicate that the sample under investigation is either isotropic or uniaxial with a $c$-plane anisotropy. As the sample rotation does not influence the MM elements, we assume that the c-axis is perpendicular to the sample surface. All further data analysis was performed using the CompleteEASE software. The datasets were combined in a multisample model with the same optical constants for each PEPI layer. We use the general oscillator approach to retrieve the dielectric permittivity function, constrained with Kramers-Kronig consistency. Typically the presence of anisotropy alters the shape of features in the ellipsometric data. In the studied case of PEPI layers, incorporation of anisotropy into the model leads to 75\% improvement of the Mean Square Error (MSE) of the dielectric permittivity function fit, which confirms that the spin-coated material exhibits anisotropy.

\medskip

\noindent{\bf Coupled oscillators model.}
     Coupled oscillator model presented by solid lines in Fig.\,\ref{rys:_FIG2_v1}b,c was fitted to the experimental data independently for measurements in horizontal and vertical polarization. 
 
For the horizontal polarization (\text{Fig.\,\ref{rys:_FIG2_v1}}b) of the polariton modes, the $4\times4$ Hamiltonian solution is: 
  \begin{equation}
  \hat{H}_{\rm H} =    \begin{pmatrix}
E_{\chi} & \Omega_{H_1}/2 & \Omega_{H_2}/2 & \Omega_{H_3}/2 \\
\Omega_{H_1}/2 & E_{\phi_{H_1}}\!\left(\textbf{k}\right) & 0 & 0\\
\Omega_{H_2}/2 & 0 & E_{\phi_{H_2}}\!\left(\textbf{k}\right)& 0 \\
\Omega_{H_3}/2 & 0 & 0 & E_{\phi_{H_3}}\!\left(\textbf{k}\right)
\end{pmatrix}.
\label{HH}
\end{equation}  
Fitting with exciton energy $E_{\chi}=2.350$\,eV resulted in coupling strengths of:
$\Omega_{H_1} = 67.2$\,meV, $\Omega_{H_2} = 88.4$\,meV, $\Omega_{H_3} = 94.4$\,meV.

For the vertical polarization (\text{Fig.\,\ref{rys:_FIG2_v1}}c) of the polariton modes, the $3\times3$ Hamiltonian solution is: 
  \begin{equation}
  \hat{H}_{\rm V} =    \begin{pmatrix}
E_{\chi} & \Omega_{V_1}/2 & \Omega_{V_2}/2 \\
\Omega_{V_1}/2 & E_{\phi_{V_1}}\!\left(\textbf{k}\right) & 0 \\
\Omega_{V_2}/2 & 0 & E_{\phi_{V_2}}\!\left(\textbf{k}\right)
\end{pmatrix}.
\label{VV}
\end{equation}  
Fitting with the same exciton energy leads to coupling strengths of:
$\Omega_{V_1} = 78.3$\,meV, $\Omega_{V_2} = 108.7$\,meV.

Dispersion relations of uncoupled photonic modes $E_{\phi_{V/H_i}}$ are marked in \text{Fig.\,\ref{rys:_FIG2_v1}}b,c with white dashed lines.
    
     \medskip

\noindent{\bf Berreman Method.}
The spectra presented in \text{Fig.\,\ref{rys:_FIG2_v1}}e,f show the reflectance calculated for H and V incident light polarization using Berreman method \cite{Schubert_PRB1996}.

Simulated cavity consists of 2 DBR mirrors made of 5 pairs of \ch{SiO2} and \ch{TiO2} layers calculated for maximal reflectance at 530\,nm. Space between the mirrors consists of LC part and 120\,nm thick perovskite layer directly on top \ch{SiO2} layer of DBR. To match position of the cavity modes with experiment, the LC layer is separated into three parts: two interface layers described by an isotropic ordinary refractive index $n_\text{o}$ with thickness of 170\,nm and central anisotropic LC layer  with thickness of 845\,nm described by diagonal dielectric tensor with extraordinary refractive index $n_\text{e}$ along $x$ direction and $n_\text{o}$ for two remaining directions. 

$n_\text{o}$ and $n_\text{e}$ values of the LC are based on Ref.\,\cite{Miszczyk_LC2018} and refractive indexes of the perovskite layer are presented in \text{Fig.\,\ref{rys:_FIG2_v1}}d.

In the simulations shown in Fig.\,\ref{rys:FIG5}c,e,k,l,m the total length of the LC cavity is equal to $L = 1362$\,nm. Twisted nematic orientation of the LC molecules around $z$ axis occurs within distances  $l = 68$\,nm at the DBR/LC interface with total 45\,deg reorientation between the top and bottom DBR. Simulations shown in Fig.\,\ref{rys:FIG5} were performed with an additional rotation of LC by 42.5\,deg around $y$ axis. Simulations were performed with 5 \ch{SiO2}/\ch{TiO2} DBR pairs centered at 555\,nm.

The dispersion of the cavity modes obtained from the Berreman method was fitted with Eq.\,\eqref{eq:Rashba2x2_2} with parameters: $E_\phi^0 = 2.0997$\,eV, $\Delta_{HV} = 13.3$\,meV, $\Delta_{AD} = -14.9$\,meV, $m_x = 1.27\times10^{-5} m_e$, $m_y = 1.38\times10^{-5} m_e$, $\delta_x = -0.32$\,meV$\mu$m$^2$, $\delta_y = -0.53$\,meV$\mu$m$^2$, $\delta_{xy} = 0.35$\,meV$\mu$m$^2$, $\alpha = -2.13\times10^{-3}$\,eV$\mu$m.

Calculations presented in Fig.\,\ref{rys:FIG6}a--f were performed for the same parameters but with $\Delta_{AD} = 18.9$\,meV observed on the sample with 30\,deg rubbing disorientation (see Supplementary Materials).

To calculate the Berry curvature we follow the procedure described in Ref.\,\cite{Bleu_PRB2018}. Based on the Stokes parameters, we define angles:  
\begin{equation}\label{eq:theta}
\Theta\!\left(\textbf{k}\right) = \arccos S_3\!\left(\textbf{k}\right),
\end{equation}
\begin{equation}\label{eq:fi}
\Psi\!\left(\textbf{k}\right) = \arctan \frac{S_2\!\left(\textbf{k}\right)}{S_1\!\left(\textbf{k}\right)},
\end{equation}
and calculate Berry curvature as
\begin{equation}\label{eq:Bz}
B_{z} = \frac{1}{2} \sin \Theta \left( \partial_{k_x} \Theta \partial_{k_y} \Psi -\partial_{k_y} \Theta \partial_{k_x} \Psi\right).
\end{equation}

	\bigskip
	
\noindent {\bf\large Acknowledgments}\\
\noindent This work was supported by the National Science Centre grants 2019/35/B/ST3/04147, 2019/33/B/ST5/02658, 2018/31/N/ST3/03046 and 2017/27/B/ST3/00271. We acknowledge the European Union's Horizon 2020 program through a FET Open research and innovation action under the grant agreement No. 899141 (PoLLoC) and No. 964770 (TopoLight). We also acknowledge the NAWA Canaletto grant PPN/BIT/2021/1/00124/U/00001. H.S. acknowledges the Icelandic Research Fund (Rannis), grant No. 217631-051.\\

\end{document}